\documentclass[journal]{IEEEtran}

\usepackage{amsmath,amssymb,latexsym,hhline,graphicx,enumerate,cite,psfrag}
\usepackage{subfigure}
\usepackage{color}
\usepackage{authblk}
\usepackage{algpseudocode}

\newtheorem{ex}{Example}

\newtheorem{rem}{Remark}

\begin{document}

\title{Low-Latency Data Sharing in Erasure Multi-Way Relay Channels}
\author[*]{Moslem Noori}
\author[**]{Hossein Bagheri}
\author[**]{Masoud Ardakani}
\affil[*]{Department of Electrical and Computer Engineering, University of British Columbia \authorcr Email: moslem@ece.ubc.ca}
\affil[**]{Department of Electrical and Computer Engineering, University of Alberta \authorcr Email: \{hbagheri, ardakani\}@ualberta.ca}

\vspace{-1cm}

\maketitle

\begin{abstract}
We consider an erasure multi-way relay channel (EMWRC) in which several users share their data through a relay over erasure links. Assuming no feedback channel between the users and the relay, we first identify the challenges for designing a data sharing scheme over an EMWRC. Then, to overcome these challenges, we propose practical low-latency and low-complexity data sharing schemes based on fountain coding. Later, we introduce the notion of end-to-end erasure rate (EEER) and analytically derive it for the proposed schemes. EEER is then used to calculate the achievable rate and transmission overhead of the proposed schemes. Using EEER and computer simulations, the achievable rates and transmission overhead of our proposed schemes are compared with the ones of one-way relaying. This comparison implies that when the number of users and the channel erasure rates are not large, our proposed schemes outperform one-way relaying. We also find an upper bound on the achievable rates of EMWRC and observe that depending on the number of users and channel erasure rates, our proposed solutions can perform very close to this bound.
\end{abstract}

\begin{keywords}
Erasure multi-way relay channels, data sharing, fountain coding, transmission strategy.
\end{keywords}

\section{Introduction} \label{Section Introduction}
\subsection{Background and Motivation}
The concept of two-way communication was first investigated by Shannon \cite{shannon_twoway} and later, multi-way channels were considered \cite{Mulen_Multi_way_IT1977}. Also, relay channels have been a prominent topic in communication theory since its early stage \cite{Meulen_Relay,Cover_Relay_TIT}. However, the combination of multi-way channels and relay channels appeared many years later in the form of two-way relay channels and multi-way relay channels (MWRCs) \cite{Rankov_Two_way_Asilomar2005,Popovski_TwoWay_ICC2006,Gunduz_Multi_way_ISIT2009}. In an MWRC, multiple users want to exchange their data with each other. The users do not have direct links to one another and a relay is used to enable the communication between them. Using the relay, data sharing between the users happen in the form of uplink (multiple-access) and downlink (broadcast) phases. Some practical examples of multi-way relaying are file sharing between several wireless devices, device-to-device communications, or conference calls in a cellular network.

MWRCs have been initially proposed and studied for Gaussian \cite{Gunduz_Multi_way_ISIT2009,Ong_FDF_Gaussian_ISIT2010} and binary symmetric \cite{Ong_FDF_CommLetter_2010} channels when the channel state information is known at the relay as well as users. Hence, they can use this information to apply appropriate channel coding. However, the channel state information may not be always known, e.g. when the links between the users and relay are time-varying. Under this situation, channel coding fails to provide reliable communication. As a consequence, the communication channel is seen as an erasure channel from the viewpoint of higher network layers where the data (packet) is received either perfectly or completely erased. Another possible situation where the erasure channel fits is a fading environment when one or more users experience a deep fade resulting in the signal loss at the relay. For more information on the erasure models for multi-user relay communication the reader is referred to \cite{Khisti_ModuluSum_Erasure_2012,Hern_ErasureTWRC_2012,Vishwanath_ITW_07}. In this work, we focus on erasure MWRCs (EMWRCs) and seek effective data sharing schemes for them.

Packet retransmission protocols are a simple solution to combat erasure. However, these protocols are wasteful in EMWRCs especially in the broadcast phase. To be more specific, if any user misses a broadcast message, a retransmission protocol forces the relay to broadcast its message to all users again. Further, implementing packet retransmission schemes or fixed-rate codes to combat erasure requires having feedback channels between the users and the relay \cite{MacKay_book_03} to carry the channel state information or acknowledgment messages for received packets. Having such feedback channels is not always feasible. Fountain coding (e.g LT codes \cite{Luby_LT_2002} or Raptor codes \cite{Shokrollahi_IT_06}) is another well-known solution which is shown to be near-optimal for erasure channels without the need for a feedback channel \cite{MacKay_book_03}. Note that by feedback channel, here, we refer to a separate communication channel used for continuously reporting channel state information or individual acknowledgment messages for received data packets to the transmitter. Even for fountain coding schemes, a final feedback message should be sent to the transmitter acknowledging the reception of all data packets. Considering the benefits of fountain coding in broadcast scenarios, in this work, we use fountain coding to develop data sharing schemes for EMWRCs. As we discuss later, implementing fountain coding for EMWRCs has many challenges. These challenges are identified and considered in the design of our strategies. It is worth noting that our work is the first one discussing EMWRCs and is different from previous studies on multi-user cooperative communications, e.g. \cite{Katti,Kliewer_ITW_2007, Koller_MWR_ISIT}, in the communication setup.

\subsection{Existing Results and Our Contributions}

The notion of fountain coding for wireless relay networks has been originally proposed in \cite{Castura_Rateless_ISIT} where one source sends its data through one or more relays to a destination. It is shown that the presented fountain coding scheme is simultaneously efficient in rate and robust against erasure. In \cite{Puducheri_DLT_IT2007}, a distributed fountain coding approach is suggested for two cases where two or four users communicate to a destination via a relay over erasure channels.  Liau \emph{et al.} \cite{Liau_LT_TCOM2011} propose a new rateless coding protocol which employs network coding and fountain coding to transmit data for Y-network. Furthermore, the work reported in \cite{Talari_LT_TCOM_2012} suggests a distributed fountain code to provide unequal error protection for two disjoint sources in a Y-network. Also, fountain coding can be exploited to relay data across multiple nodes \cite{Gummadi_RelayFountain_ITW2008} or data broadcast \cite{Cataldi_LT_2010} in a network. 

In addition, \cite{Molisch_Rateless_Globecom_2006,Gong_Rateless_TWireless2012,Uppal_Rateless_TSP2011} consider fountain coding scenarios for different setups of relay networks over fading channels. Molisch \emph{et al.} consider a cooperative setup in \cite{Molisch_Rateless_Globecom_2006} where one source sends its data to a destination through multiple relays and argue that using fountain coding reduces the energy consumption for data transmission from the source to the destination. Also, in a fading environment, \cite{Gong_Rateless_TWireless2012} and \cite{Uppal_Rateless_TSP2011} apply fountain coding to improve the performance in a four-node (two sources, one relay, and one destination) and a three-node (one source, one relay, and one destination) setup respectively.

Applying fountain coding to EMWRCs, however, has its own challenges. First, it is undesirable to perform fountain decoding and re-encoding at the relay as it requires waiting for all data packets of all users and needs extra hardware at the relay. To avoid this latency and also decrease the implementation cost, we are interested in data sharing solutions that can work with fountain coding/decoding only at the users. Second, each user needs to simultaneously track the combinations of packets formed at all other users. This needs to be accomplished without adding significant overhead or hardware complexity to the system. Third, since data of all users are mixed during the transmission, in the case of using the popular belief propagation decoders, fountain decoding will almost surely fail at some users as the received degree distribution will differ from that of the transmitted one. In particular, the weight of degree-one equations will be very small (due to mixing at the relay). Since degree-one equations play a key role in fountain decoding, this can cause the decoder to stop at early stages. Thus, the users' data sharing strategies must be designed to combat this problem. Finally, we like to have data sharing strategies that are readily scalable with the number of users.

It is important to notice that the existence of the side information in each user (i.e. each user knows its own data) makes EMWRCs different from one-way relay networks in which a set of users, called sources, send their data to another set, called destinations. An efficient data-sharing strategy for EMWRCs should make use of this side information effectively.

The focus of this paper is on devising efficient data sharing strategies based on fountain codes for EMWRCs. Considering the design challenges pointed above, we devise two data sharing schemes that (i) need fountain coding/decoding only at the users' side (thus they have low latency) (ii) work with synchronized fountain encoders (hence, does not expose extra overhead or hardware complexity) (iii) can decode each user's data separately (thus fountain decoding will not fail) and (iv) are easily scalable with the number of users. We also show that the system's performance can be further improved by performing simple matrix operations at the relay as well as shuffling the users' transmission order.

To evaluate the performance of the proposed schemes, we introduce the concept of \emph{end-to-end erasure rate} (EEER). Using EEER, we compare the achievable rates of our schemes with the existing conventional one-way relaying (OWR). Furthermore, we derive an upper bound on the achievable data rates of the considered EMWRC. The achievable rates of our schemes are then compared with this bound to determine their performance gap. This comparison reveals that depending on the uplink and downlink erasure probabilities and number of users, our proposed data sharing strategies can get very close to the rate upper bound and outperform OWR. The proposed schemes are also compared with OWR in terms of their transmission overhead. The implication of this comparison is that for small erasure probabilities or small number of users, the proposed schemes accomplish data sharing between users with a smaller overhead than OWR.

\section{System Model} \label{Section System model}
In this paper, we study an EMWRC with $N$ users, namely $u_1,u_2,\ldots,u_N$. The users want to fully exchange their information packets. Users do not have direct link between themselves and thus the communications happens with the help of a (low-complexity) relay. Each user has $K$ information packets and we assume that the information packets are seen as data bits. It means that for the $k$th packet at $u_i$, denoted by $m_{i,k}$, we have $m_{i,k} \in \{ 0,1 \}$. Also, at a given transmission turn, the transmit message of $u_i$, derived from its information messages $m_{i,1}, \ldots, m_{i,K}$, is denoted by $x_i \in \{0,1 \}$. Although the channel inputs are binary, the channel outputs are from a ternary alphabet $\{0,1,E\}$. Here, $E$ denotes the erasure output.

The general communication model for such a system can be described as follows. To share their data, users first send their transmit messages in the uplink phase. In each uplink phase, some (or all) users send their data to the relay. The transmitted packet of $u_i$ experiences erasure with probability $\epsilon_{u_i}$ in the uplink phase. Here, we define a  Bernoulli variable $b_i$ representing the state of $x_i$ in the uplink transmission. More specifically, $b_i = 1$ (with probability $1 - \epsilon_{u_i}$) means that $x_i$ has not been erased in the uplink and $b_i = 0$ indicates an erasure. Using this variable, the received signal at the relay can be modeled as
\begin{equation}\label{Eq uplink received model}
y_{\mathrm{r}} = \left\{ \begin{array}{ll}
E & \text{if for all }  i = 1,2,\ldots,N : b_i = 0   \\
\displaystyle{\bigoplus_{i=1}^{N}} a_{i} \, b_{i}  \, x_{i} &\mbox{otherwise}
\end{array} \right.
\end{equation}
where the summation is a modulo-2 sum. In (\ref{Eq uplink received model}), $a_i$ is a binary variable showing whether $x_i$ is transmitted in the uplink or not. For $u_i$, $a_i = 1$ indicates that $x_i$ is transmitted and $a_i = 0$ otherwise. 

A similar transmission model has been considered in \cite{Hern_ErasureTWRC_2012, Khisti_ModuluSum_Erasure_2012} to model erasure two-way relay and multiple-access channels. The model in (\ref{Eq uplink received model}) mimics a wireless multiple-access channel where users transmit their data over a fading environment \cite{Hern_ErasureTWRC_2012}. When some users go into the deep fade, the relay loses their signal and their transmitted data are erased. In the case of deep fade over all users, the relay does not receive a meaningful signal which is seen as an erasure. The interested readers are referred to \cite{Hern_ErasureTWRC_2012} for more information on the considered model for erasure channels. Note that here, the data combining, i.e. summation of the transmitted packets, is done by the channel which is mainly referred to as \emph{physical-layer network coding} \cite{Wu_PNC_TCOM2005}.

After receiving the users' data in the uplink phase, the relay forms its message $x_{\mathrm{r}}$ based on $y_{\mathrm{r}}$. In the downlink, relay broadcasts its message to all users. $u_i$ misses relay's broadcast message with erasure probability $\epsilon_{d_i}$ and receives it with probability $1 - \epsilon_{d_i}$. 

After receiving the relay's broadcast message, each user first tries to separate different users' data from each other and then decodes them. The uplink and downlink transmissions should continue until each user is able to retrieve the information packets of any other user (full data exchange). 

\section{Data Sharing Schemes} \label{Section Schemes}
In this section, we propose our data sharing schemes for the discussed EMWRCs and present their specific communication model. Our proposed data sharing schemes consist of four principal parts: i) Fountain coding at the users, ii) Users' transmission strategy, iii) Relay's transmission strategy, and iv) Data separation at the users. In the rest of this section, we discuss each of these parts in details. The performance gap of these schemes is later evaluated by comparing their achievable common data rates with a rate upper bound derived in Section \ref{Section Upper bound}. Here, the achievable common data rate refers to a data rate that all users can reliably exchange their data with this rate.

\subsection{Fountain Coding}
To sustain reliable communications in an EMWRC, an appropriate scheme should be employed to combat erasure. Retransmission protocols are a simple approach for this purpose, however, they are wasteful for EMWRCs due to the significantly large number of transmissions that is needed to ensure receiving data by all users in the BC phase \cite{MacKay_book_03}. Furthermore, implementing retransmission protocols as well as conventional fixed-rate erasure correcting codes (e.g. Reed-Solomon codes) requires a feedback channel between the users and the relay carrying  acknowledgment messages or channel state information. Another approach for combating data erasure is fountain coding which provides reliable data communication without the need for prior information of the channel state at the transmitter. In the following, we describe how fountain coding is employed in our proposed data sharing schemes.

If relay wants to perform fountain decoding and re-encoding before forwarding the data to the users, it should wait to receive all data packets from all users and then decode them. This causes a significant delay in the data sharing process. Thus, in our proposed solution, the fountain encoding and decoding are performed only at the users. More specifically, $u_i$ encodes its information packets, $m_{i,k}$ where $k = 1,2,\ldots,K$, with a fountain (e.g. a Raptor \cite{Shokrollahi_IT_06}) code and forms its transmit message $x_i$. As mentioned previously, we denote the packets by binary symbols for the sake of simplicity.

Here, it is assumed that the fountain encoders at the users are synchronized. With synchronized encoders, each user can easily keep track of the combinations of the packets formed at the other users without exposing extra hardware complexity or overhead to the system. Knowing the combination of the formed packets is important to proceed with the fountain decoding at the users. To implement synchronized fountain encoders, users have identical random number generators with equal initial seeds. The initial seed can be passed to the users from the relay at the beginning of the communication. This is a one-time setup and does not need to be changed during the data communication. If a new user joins the data communication, it can simply obtain the proper initial seed for its random number generator from the relay. Alternatives to our synchronized encoder assumption could be implemented by relaxing the overhead or hardware complexity constraint. For instance, users can include information about their packet combination in the overhead of the transmitted packets. They also can use $N - 1$ independent random number generators, fed with proper initial seeds, to keep track of other users packet combination. Any of the mentioned methods can be employed in our proposed schemes to provide to the packet combinations at the decoders without changing the decoding performance performance.

After encoding their packets, users send them in the uplink phase. They continue transmitting fountain-coded packets until the data sharing is finished and all users have the full data of any other user. When a user receives the data completely, it sends a 1-bit acknowledgment message to the relay. After receiving acknowledgment message from all users, relay broadcasts a 1-bit message to all users announcing that the data communication is over, so the users stop transmitting packets.  This 1-bit final feedback is common in practice and makes sure that all data has been carried successfully and prevents immature data transmission.

Assuming $K$ information packets at each user, if data sharing is accomplished after sending the $K'$th encoded packet, the overhead is defined as $O = \frac{K' - K}{K}$ \cite{MacKay_book_03}. Please note that here, we consider the transmission overhead to evaluate the performance of the data sharing strategies. Another commonly-used measure for fountain codes is the reception overhead which depends on the characteristics of the underlying fountain code. Since we do not deal with the fountain code design, reception overhead is irrelevant to our discussions. 

\subsection{Users' Transmission Strategies}
In our proposed data sharing schemes, we define a \emph{round of communication} consisting of $L$ uplink and $L$ downlink transmissions (time slots). During one round of communication, users want to exchange one of their fountain coded packets. Depending on the users' transmission strategy, a set of users simultaneously send their fountain coded packets to the relay in each of these $L$ time slots. A users' transmission strategy is determined by the transmission matrix $\textbf{A} = [a_{l,i}]_{L \times N}$. According to $\textbf{A}$, $u_i$ transmits in $l$th uplink slot if $a_{l,i} = 1$. Otherwise, $u_i$ stays silent and does not transmit.

In the $l$th uplink slot, the relay's received signal is
\begin{align}\label{Eq relay signal}
y_{\mathrm{r},l}=\bigoplus_{i=1}^{N} a_{l,i} \, b_{l,i} \, x_{i}.
\end{align}
In (\ref{Eq relay signal}), $b_{l,i}$ is a Bernoulli random variable representing the erasure status of $x_i$ in the $l$th uplink slot. Here, $b_{l,i} = 0$ with probability $\epsilon_{u_i}$ and $b_{l,i} = 1$ with probability $1 - \epsilon_{u_i}$. Defining $\textbf{x} = [x_i]_{N \times 1}$ and $\textbf{y}_\mathrm{r} = [y_{\mathrm{r},l}]_{L \times 1}$, (\ref{Eq relay signal}) can be rewritten in the following matrix form
\begin{equation} \label{Eq relay signal matrix}
\textbf{y}_{\mathrm{r}} = (\textbf{A} \odot \textbf{B}) \textbf{x} = \textbf{A}_{\mathrm{r}} \textbf{x}.
\end{equation}
In (\ref{Eq relay signal matrix}), $\textbf{B} = [b_{l,i}]_{L \times N}$ and $\odot$ represents the Hadamard product. Also, $\textbf{A}_{\mathrm{r}}$ is the relay's received matrix. Please note that according to (\ref{Eq uplink received model}), if all $b_{l,i}$'s are 0 in an uplink transmission slot, its associated element in $\textbf{y}_{\mathrm{r}}$ is $E$.

In this work, we consider three different users' transmission strategies: conventional one-way relaying and our proposed pairwise transmission strategies.

\subsubsection{One-Way Relaying (OWR)}
In this scheme, $L\!=\!N$, and the data of each user is solely sent to the relay in one of the uplink slots. For OWR, the uplink transmission matrix $\textbf{A}$ is an $N\!\times\!N$ identity matrix, i.e. $\textbf{A}=\text{I}(N)$.

\subsubsection{Minimal Pairwise Relaying (MPWR)}
The scheme divides the uplink and downlink into $L\!=\!N\!-\!1$ transmissions. A sequential pairwise data communication to the relay is used in MPWR. In particular, in time slot $l$ of the uplink, $u_l$ and $u_{l\!+\!1}$ transmit to the relay. The pairwise scheme is shown to be capacity achieving when the links are binary symmetric \cite{Ong_FDF_CommLetter_2010}. The MPWR's uplink transmission matrix is
\begin{align}
\textbf{A}&=
\left( \begin{array}{ccccccc}
1 & 1 & 0 & 0 & \ldots & 0 & 0 \\
0 & 1 & 1 & 0 & \ldots & 0 & 0 \\
\vdots\\
0 & \ldots &  &  &  & 1 & 1
\end{array} \right)_{(N-1)\times N}.
\end{align}

\subsubsection{One-Level Protected Pairwise Relaying (OPPWR)}
By using one extra uplink time slot compared to MPWR and sending a pairwise combination of the first and the last users, OPPWR has an extra protection against erasure compared to MPWR. More specifically, it can tolerate at least one erasure either in the uplink or in the downlink transmissions, which does not hold for the MPWR scheme. For this scheme,
\begin{align}
\textbf{A}&=
\left( \begin{array}{ccccccc}
1 & 1 & 0 & 0 & \ldots & 0 & 0 \\
0 & 1 & 1 & 0 & \ldots & 0 & 0 \\
\vdots\\
0 & \ldots &  &  &  & 1 & 1\\
1 & 0 & \ldots &  &  & 0 & 1
\end{array} \right)_{N\times N}.
\end{align}

\begin{figure}[h!]
\centering
\psfrag{a1}{$u_1$}
\psfrag{a2}{$u_2$}
\psfrag{a3}{$u_{i}$}
\psfrag{a4}{$u_{(i+1)}$}
\psfrag{a5}{$u_{(\!N\!-\!1\!)}$}
\psfrag{a6}{$u_{N}$}
\psfrag{x1}{$x_i$}
\psfrag{x2}{$x_{i+1}$}
\subfigure[$i$th MAC phase]{
    \label{Fig:MAC}
   \includegraphics[scale =.11]{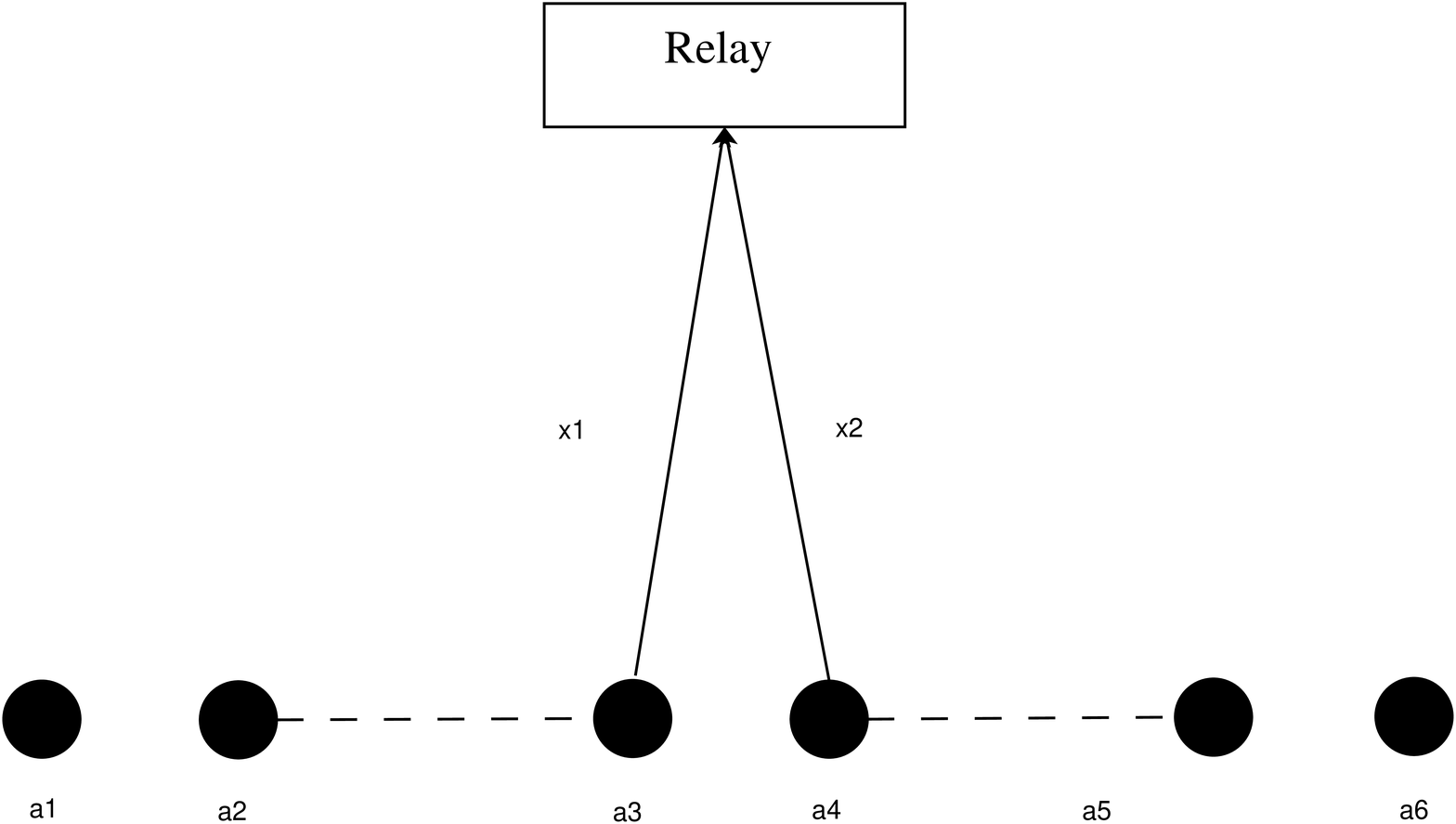}
 }  
\psfrag{a1}{$u_1$}
\psfrag{a2}{$u_2$}
\psfrag{a3}{$u_{i}$}
\psfrag{a4}{$u_{(i+1)}$}
\psfrag{a5}{$u_{(\!N\!-\!1\!)}$}
\psfrag{a6}{$u_{N}$}
\psfrag{y1}{$x_i \oplus x_{i+1}$}
 \vspace{-.2cm} \subfigure[$i$th BC phase]{
    \label{Fig:BC}
   \includegraphics[scale =.11] {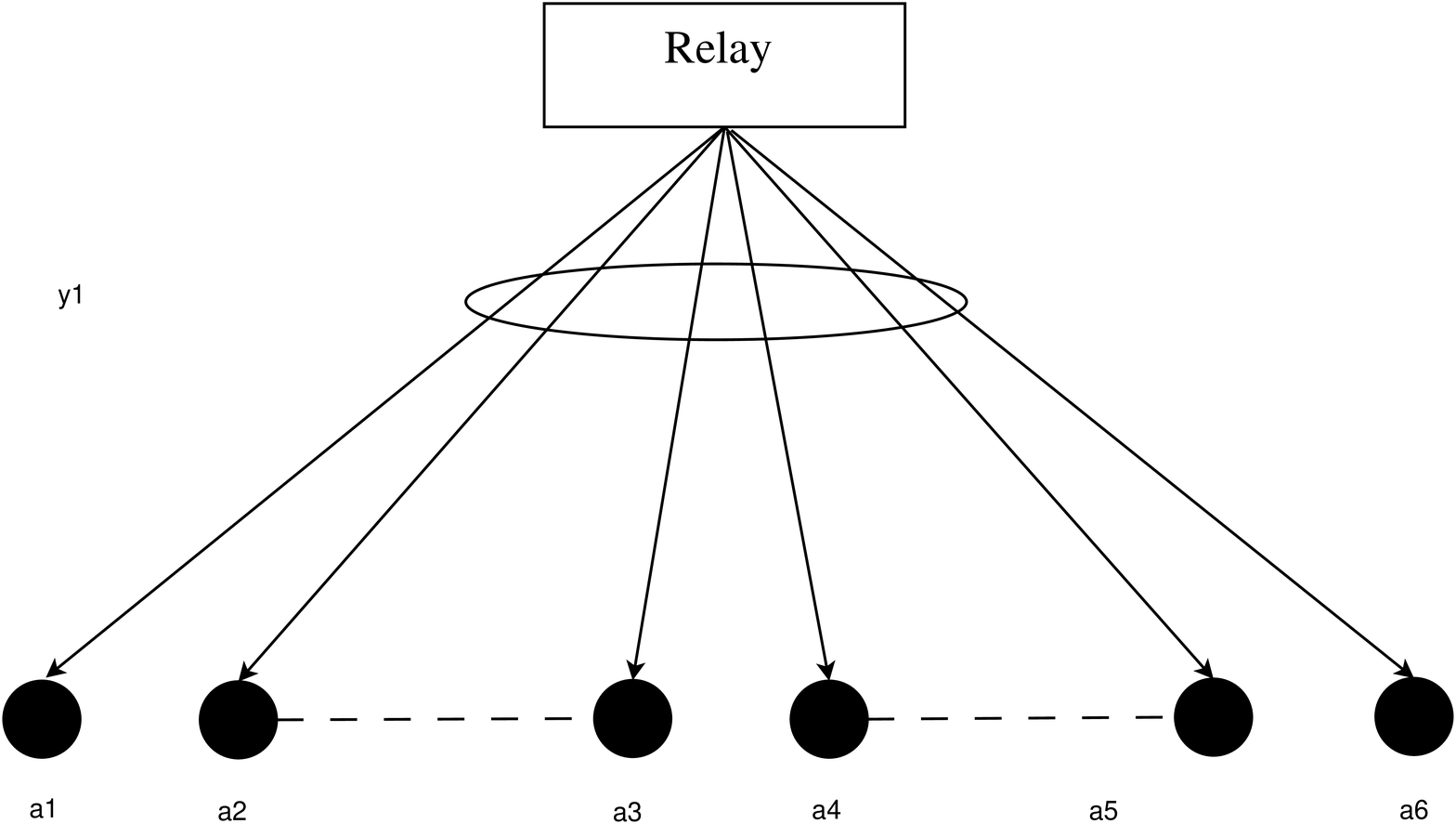}
 }
\label{Fig Visual Presentation}
\caption{Demonstration of the $i$th MAC and BC phases}
\end{figure}

Figure~\ref{Fig:MAC} depicts the uplink transmission for the fountain coded packets of $u_i$ and $u_{i +1}$ assuming. The downlink broadcast phase is presented in Figure~\ref{Fig:BC}.

Note that $\textbf{A}$ can be seen as a simple outer code at the network level. This outer code provides protection against the users' packet loss. While  MPWR supports correcting one uplink erasure except for the first and last transmitting users, OPPWR can guarantee one level of protection in both uplink and downlink. This extra level of protection comes at the cost of one additional uplink and downlink slots resulting in a rate degradation by a factor of $\frac{N - 1}{N}$. Extra protection on the packets can be obtained by changing the transmission strategy through another design for $\textbf{A}$. However, this may result in a more complex structure for $\textbf{A}$, and consequently less scalable strategy, or extra transmissions in the uplink and downlink resulting in a rate loss.

\subsection{Relay's Transmission Strategy}
After receiving $\textbf{y}_{\mathrm{r}}$ in the uplink phase, relay forms its message $\textbf{x}_{\mathrm{r}} = [x_{\mathrm{r},l}]_{L \times 1}$ based on $\textbf{y}_{\mathrm{r}}$. Then, $\textbf{x}_{\mathrm{r}}$ is sent to the users in $L$ downlink transmissions. As mentioned before, we like to sustain a low-latency and simple relaying. To this end, we consider two different scenarios for the relay to form its message, $\textbf{x}_{\mathrm{r}}$.

In the first scenario, relay simply forwards its received signal, i.e. $x_{\mathrm{r},l} = y_{\mathrm{r},l}$, in each time slot. In this case, relay does not need to buffer the received signals in the uplink slots and has the minimum relaying latency.

In the second case, users first transmit their data in $L$ consecutive uplink transmissions. Here, relay has a buffer with length $L$ for its received signal and is capable of performing simple elementary matrix operations. By buffering the received signals in the uplink slots and knowing which packets have been erased\footnote{This knowledge can be provided to the relay by adding orthogonal $N$-tuple of bits at the header of the users' packets. To this end, $u_i$ sets the $i$th bit of the tuple to 1 and the rest to 0. Relay can identify which user's packet is erased when its associated bit is not 1. Furthermore, in a practical wireless system, relay may obtain this knowledge through pilot channels or employing successive interference cancellation.}, the relay forms $\textbf{A}_{\mathrm{r}}$. Now, in the case of erasure events in the uplink, relay performs elementary matrix operations on $\textbf{A}_{\mathrm{r}}$ and tries to retrieve all erased elements of the original transmitted matrix $\textbf{A}$ or at least some of them. The result of the matrix operations on $\textbf{A}_{\mathrm{r}}$ is called $\tilde{\textbf{A}}$. Relay then performs the same matrix operations on $\textbf{y}_{\mathrm{r}}$ to form $\textbf{x}_{\mathrm{r}}$. In other words, $\textbf{x}_{\mathrm{r}} = \tilde{\textbf{A}} \textbf{x}$. We call this method \emph{matrix reconstruction}. The elements of $\tilde{\textbf{A}}$ are basically equal to $\textbf{A}$ except those elements that are erased in the uplink and cannot be retrieved through matrix reconstruction. Since relay may be able to retrieve some of the erased elements of $\textbf{A}$, doing matrix reconstruction can lower the effective uplink erasure rate. Then, relay broadcasts its messages in $L$ downlink slots to the users. While the relay does not perform any fountain decoding/encoding here, it may try to decode outer code presented by $\textbf{A}$ via matrix reconstruction. Matrix reconstruction is a simple and fast operation and since in practice $L$ is much smaller than $K$, the low-latency requirement is still met.

\ex Consider an EMWRC with $N = 3$ users and OPPWR is used as the users' transmission strategy. In this case,
\begin{equation}
\textbf{A} = \left( \begin{array}{ccc}
1 & 1 & 0 \\
0 & 1 & 1 \\
1 & 0 & 1
\end{array} \right).
\end{equation}
Now, assume that in the third uplink slot, $u_3$'s data has been erased. Thus
\begin{equation}
\textbf{A}_{\mathrm{r}} = \left( \begin{array}{ccc}
1 & 1 & 0 \\
0 & 1 & 1 \\
1 & 0 & 0
\end{array} \right).
\end{equation}
If the relay does the modulo-2 sum of the first and second rows of $\textbf{A}_{\mathrm{r}}$, it can retrieve $\textbf{A}$. Thus, in this case $\tilde{\textbf{A}} = \textbf{A}$. Note that if the relay does not perform reconstruction and $x_{\mathrm{r},2}$ is erased in the downlink, $x_3$ will be lost, but with reconstruction, it can be retrieved. Note that if MPWR is applied for the same scenario, relay cannot fix the affected equation and $x_3$ is not recoverable.

\subsection{Data Separation}
After receiving the downlink signal from the relay and knowing its own transmitted packet, each user first separates the data of other users before proceeding with the fountain decoding. After separating data packets, the user buffers them to proceed with the fountain decoding\footnote{Here, it is assumed that the users know matrix $\tilde{\textbf{A}}$. This can be achieved in practice by adding $N$ bits to the header of each packet. In practical cases, this extra overhead is negligible compared to the size of the packets.}. If the data separation is not done, the user should treat all data from all other users as a large stream of fountain coded packets. This can result in the failure of fountain decoding due to not receiving enough degree-one packets. Here, degree-one packets refer to the fountain-coded packets which are composed at the encoder from only one original data packet. This type of packets play an important role in fountain decoding and helps the decoder to start and continue decoding.

Let $\textbf{y}_i=[y_{l,i}]_{L \times 1}$ be the received vector at $u_i$ after one round of communication. Here, either $y_{l,i} = x_{\mathrm{r},l}$ or $y_{l,i} = E$. The received downlink signal at $u_i$ can be written in the following matrix form
\begin{equation}
\textbf{y}_i = \textbf{A}_{\mathrm{r}_i} \textbf{x}
\end{equation}
where $\textbf{A}_{\mathrm{r}_i}$ is the received matrix at $u_i$. Here, the rows of $\textbf{A}_{\mathrm{r}_i}$ are equal to the rows of $\tilde{\textbf{A}}$ except that some rows are erased.

Without loss of generality, we consider the data separation at $u_1$. Knowing its own data packet, $u_1$ tries to find other users' transmitted data by solving the following system of linear equations
\begin{equation} \label{Eq linear system}
\textbf{A}_1 \textbf{x} = [x_1 \, \textbf{y}_1^T]^{T}
\end{equation}
where
\begin{equation}\label{Eq A_1}
\textbf{A}_1 = \left( \begin{array}{ccccc}
1 & 0 & \ldots & 0 & 0 \\
&  & \textbf{A}_{\mathrm{r}_1} & &
\end{array} \right).
\end{equation}
The transmitted packet of user $j$, $x_j$, is erased at $u_1$ when it cannot be retrieved by solving (\ref{Eq linear system}). From (\ref{Eq A_1}), it is seen that $L$ should be at least $N - 1$ to make data separation feasible. After separating the data packets of each user, $u_1$ waits until receiving enough packets to proceed with the fountain decoding.

To further clarify our proposed schemes, the overall data sharing process based on our proposed schemes is summarized in Figure~\ref{Algorithm}.

\begin{figure}
\begin{algorithmic}[1]
\Repeat 
\State Consider $L$ uplink and downlink slots
\State Form fountain coded packet $x_i$ at $u_i$
\For {$i=1:L$}
\State Users transmit based on $i$th row of \textbf{A}
\If{no reconstruction}
\State Relay simply forwards $y_{\mathrm{r},l}$
\Else
\State buffer $y_{\mathrm{r},l}$.
\If{$i = L$}
\State Relay forms $\textbf{A}_{\mathrm{r}}$
\State Relay finds $\tilde{\textbf{A}}$ by matrix operations over $\textbf{A}_{\mathrm{r}}$
\State Relay broadcasts messages based on $\tilde{\textbf{A}}$
\EndIf
\EndIf
\EndFor
\State Users perform data separation.
\State Users do fountain decoding.
\Until{All users receive the data}
\end{algorithmic}
\caption{Algorithmic presentation of the proposed schemes.}\label{Algorithm}
\end{figure}

\ex Consider an EMWRC with $N = 4$ users. In this EMWRC, MPWR is used and the relay simply forwards its received messages without doing reconstruction. In this case,
\begin{equation}
\textbf{A} =
\left( \begin{array}{cccc}
1 & 1 & 0 & 0 \\
0 & 1 & 1 & 0 \\
0 & 0 & 1 & 1
\end{array} \right).
\end{equation}
Now, assume that $x_2$ is erased in the second uplink transmission. Also, $x_{\mathrm{r},3}$ has been erased in the downlink and the received signal at $u_1$ is $\textbf{y}_1 = [0 \, 1 \, E]^T$. Assuming $x_1 = 1$, $u_1$ forms the following system of linear equations to find $x_2$, $x_3$ and $x_4$:
\begin{equation}\label{Eq linear system example}
\left( \begin{array}{cccc}
1 & 0 & 0 & 0 \\
1 & 1 & 0 & 0 \\
0 & 0 & 1 & 0 \\
0 & 0 & 0 & 0
\end{array} \right)
\left( \begin{array}{c}
x_1 \\
x_2 \\
x_3 \\
x_4
\end{array} \right) =
\left( \begin{array}{c}
1 \\
0 \\
1 \\
E
\end{array} \right).
\end{equation}
From (\ref{Eq linear system example}), $u_1$ finds that $x_2 = x_3 =  1$ while $x_4$ is declared as erasure.  

In our proposed schemes, we have assumed that the transmitter (e.g users in the uplink and relay in the downlink) does not have any knowledge about the channel state before sending data. The availability of such information through adding a feedback channel can be used to coordinate communication in the system for possible improvements.

\section{End-to-end Erasure Rate} \label{Section EEER}
To study the performance of the three aformentioned schemes, we introduce a useful concept called end-to-end erasure rate (EEER). This concept is helpful in: i) finding the achievable rates of the schemes, and ii) calculating their transmission overhead.

Consider an arbitrary user, $u_i$. For any $j \neq i$, if we are able to identify the erasure rate of $u_j$'s packets at $u_i$, denoted by $\epsilon_{i,j}$, we can simply model the communication between this pair of users with an erasure channel with the erasure probability of $\epsilon_{i,j}$. The achievable data rate over this channel is then $1 - \epsilon_{i,j}$. Also, the transmission overhead of an ideal fountain code for data transmission from $u_j$ to $u_i$ over this channel is
\begin{equation}\label{Eq Overhead u_i to u_j}
O_{i,j} = \frac{\epsilon_{i,j}}{1 - \epsilon_{i,j}}.
\end{equation}
Based on the above discussion, we define pairwise EEER which is the erasure rate between a pair of users where one of them serves as the data source and the other one as destination. Having $N$ users in the systems results in $\frac{N(N-1)}{2}$ pairwise EEERs. Now, we define maximum EEER, which we simply call EEER and denote it by $\epsilon_f$, as the maximum erasure rate over all pairs of users. In other words, $\displaystyle{\epsilon_f = \max_{i,j} \epsilon_{i,j}}$. Since the achievable common data rate, $R$, is determined by the data transmission rate between the users experiencing the worst erasure, we have $R = 1 - \epsilon_{f}$. With a similar argument, the overall transmission overhead is
\begin{equation}\label{Eq Overhead using EEER}
O = \frac{\epsilon_f}{1 - \epsilon_f}.
\end{equation}
Please note that in practice, the transmission overhead is larger than (\ref{Eq Overhead using EEER}) due to using non-ideal fountain codes.

\subsection{EEER Calculation for OWR}
Using OWR, a packet sent from user $i$ is received by user $j$ if it is not erased neither in the uplink nor in the downlink. Thus, defining $\bar{\epsilon}_{u_i} = 1 - \epsilon_{u_i}$ and $\bar{\epsilon}_{d_j} = 1 - \epsilon_{d_j}$, we have $\epsilon_{i,j} = 1 - \bar{\epsilon}_{u_i} \bar{\epsilon}_{d_j}.$ Now, EEER is
\begin{equation}
\epsilon_{f}^{\mathrm{OWR}} = \max_{i,j} \epsilon_{i,j} = 1 - \min_{i,j} \bar{\epsilon}_{u_i} \bar{\epsilon}_{d_j}.
\end{equation}
Note that the reconstruction process at the relay is not helpful when OWR is used since the relay receives the data of a specific user in only one uplink channel use. Further, for a symmetric EMRWC where for all $i$, $\epsilon_{u_i} = \epsilon_u$ and $\epsilon_{d_i} = \epsilon_d$, pairwise EEERs are all equal for any pair of users.

\subsection{EEER Calculation for MPWR}
For MPWR, the relay receives the data of each user (except the first and the last ones) in two uplink time slots. Thus, it may be able to employ data reconstruction for $u_2$ to $u_{N - 1}$ in order to retrieve their data if it is erased in only one uplink transmission. In the following, we study EEER for both cases when the relay does not perform data reconstruction and when it does.

\textit{MPWR without Reconstruction:} First, we study $\epsilon_{i,1}$, the pairwise EEER of $u_i$, $i=2,\ldots,N$, at $u_1$. Then we extend the analysis to other users. For decoding at $u_1$, let us call the probability of finding $x_i$ at $i$th or $(i+1)$th rows of $\textbf{A}_1$ by $P_1^1(i)$ and $P_2^1(i)$ respectively.

First, we calculate $P_1^1(i)$. Notice that $P_1^1(1) = 1$ since $x_1$ is always known at $u_1$. For $i > 1$, $x_i$ is found in row $i$ when this row is not erased in the downlink phase and : (i) No erasure has happened in row $i$ during the uplink phase and the value of $x_{i-1}$ has been found from row $i - 1$ or (ii) In the $i$th row, $x_{i-1}$ was erased in the uplink phase, while $x_i$ has been perfectly received (only $x_i$ exists in this row). Hence,
\begin{equation}\label{Eq P_1}
P_1^1(i) = \bar{\epsilon}_{d_1} (\bar{\epsilon}_{u_i} \bar{\epsilon}_{u_{i - 1}} P_1(i-1) + \bar{\epsilon}_{u_{i}} \epsilon_{u_{i-1}}).
\end{equation}
Having $P_1^1(1) = 1$, by solving the above recursive equation for $i = 2,\ldots,N$, all $P_1^1(i)$'s are found.

Now, we calculate $P_2^1(i)$. Since $x_N$ appears just once in (\ref{Eq linear system}) when MPWR is used, $P_2^1(N) = 0$. Also $P_2^1(1) = 1$. By a logic similar to the one used for the calculation of $P_1^1(i)$, for $i=2,\ldots,N-1$, we have
\begin{equation}
P_2^1(i) = \bar{\epsilon}_{d_1} (\bar{\epsilon}_{u_i} \bar{\epsilon}_{u_{i + 1}} P_2(i+1) + \bar{\epsilon}_{u_i} \epsilon_{u_{i+1}}).
\end{equation}

Now, to complete the pairwise EEER calculation, we just need to find $P_c^1(i)$ representing the probability of finding $x_i$ at $u_1$ in both $i$ and $(i +1)$th equations. Here, $x_i$ can be retrieved from both $i$th and $(i+1)$th rows if none of these rows is erased in the downlink and $x_i$ does exist in both rows. Also, one of these situations should happen: (i) $x_{i-1}$ in row $i$ and $x_{i+1}$ in row $i+1$ are both erased in the uplink phase, (ii) Either $x_{i-1}$ or $x_{i+1}$ is erased in the uplink phase and the other one was found before solving the corresponding equation, (iii) Nothing is erased in the uplink phase and $x_{i-1}$ and $x_{i+1}$ have been previously found. Thus, for $i=2,\ldots,N$, we have
\begin{align}\label{Eq P_c}
P_c^1(i) \! & =  \bar{\epsilon}_{d_1}^2  \bar{\epsilon}_{u_i}^2 \! \Big[\epsilon_{u_{i-1}} \epsilon_{u_{i+1}}  \! + \! \epsilon_{u_{i+1}} \bar{\epsilon}_{u_{i-1}} P_1^1(i\!-\!1\!) \! \\ \nonumber
& +  \epsilon_{u_{i-1}} \bar{\epsilon}_{u_{i+1}} P_2^1(i\!+\!1\!) \!+\! \bar{\epsilon}_{u_{i+1}} \bar{\epsilon}_{u_{i+1}} P_1^1(i\!-\! 1) P_2^1(i \!+\!1\!)\Big].
\end{align}

Now, the probability of finding $x_i$ at $u_1$ is
\begin{equation}
P^1(i) = P_1^1(i) + P_2^1(i) - P_c^1(i)
\end{equation}
and $\epsilon_{i,1} = 1 - P^1(i)$.

Let us derive the probability of finding $x_i$ at user $j$, called $P^j(i)$. Since $x_j$ is known at user $j$, finding the values of $x_{j-1},x_{j-2},\ldots,x_1$ can be seen as finding $x_2,x_3,\ldots,x_j$ at $u_1$ when there are only $j$ users in the system trying to exchange their data. Thus, for $i=1,2,\ldots,j-1$, $P^j(i) = P^1(j - i +1)$ where $P^1(\cdot)$ is calculated when there are $j$ users in the system. Similarly, for $i = j+1,j+2,\ldots,N$, we have $P^j(i) = P^1(i - j + 1)$ when only $N - j + 1$ users exchange their data. Hence, $\epsilon_{i,j}$ is derived.

Similar to OWR, $\epsilon_{f}^{\mathrm{MPWR}} = \displaystyle{\max_{i,j} \epsilon_{i,j}}$. Furthermore, the average erasure rate that each user experiences is
\begin{equation} \label{Eq Average erasure MPWR}
\epsilon_{\mathrm{ave}}^{MPWR} = 1 - \frac{\displaystyle{\sum_{j=1}^N \sum_{i=1,i \neq j}^N P^j(i)}}{N (N -1)}.
\end{equation}
The importance of $\epsilon_{\mathrm{ave}}^{MPWR}$ is later discussed in Subsection \ref{Subsection Numerical Examples}.

\rem \label{Remark MPWR} Assume a symmetric EMWRC where $\epsilon_{u_i} = \epsilon_{u}$ and $\epsilon_{d_i} = \epsilon_{d}$ for all $i$. In this case, unlike OWR, pairwise EEERs are not necessarily equal when MPWR is used. Further, it can be shown that
\begin{equation}
\min_{j,i} P^j(i) = P^1(N) = P^N(1).
\end{equation}
Thus, $\epsilon_f^{\mathrm{MPWR}} = \displaystyle{\max_{i,j} \epsilon_{i,j} = 1 - P^1(N)}$.

\textit{MPWR with Reconstruction:} Reconstruction at the relay is performed on $\textbf{A}_{\mathrm{r}}$ and gives $\tilde{\textbf{A}}$. Its purpose is to reduce the uplink erasure rate without affecting the downlink. In the following, we find the equivalent uplink erasure rate when MPWR along with relay reconstruction is used. The equivalent uplink erasure probability of $x_i$ in $j$th pairwise transmission is the probability of not being able to retrieve it at $j$th equation even after reconstruction at the relay. Notice that $x_i$ appears in $(i-1)$th and $i$th equations of $\textbf{A}$. Thus, $j \in \{i-1 ,i \}$.

First of all, if $x_1$ or $x_N$ is erased in its associated transmission, it never can be retrieved since these data packets appear in only one row of $\textbf{A}$. Now, assume that $x_i$, $2 \leq i \leq N - 1$, is erased in $(i - 1)$th equation. To find $x_i$ from the rest of equations, one of these cases should happen: i) $x_{i+1}$ is erased in $i$th equation while $x_i$ exists there, ii) Both $x_i$ and $x_{i+1}$ exist in $i$th equation, and only $x_{i+1}$ is received by the relay in $(i+1)$th equation, and so on. This continues until the case where all $x_i$'s in the $i$th to $(N-2)$th equations exist and $x_N$ is erased from the $(N-1)$th row of $\textbf{A}$ while $x_{N-1}$ exists. Thus, the probability of retrieving $x_i$ in the $(i-1)$th row of $\textbf{A}_{\mathrm{r}}$ when it has been originally erased in the uplink transmission is
\begin{align} \nonumber
P_{c}^{i,i-1}  \! & = \! \bar{\epsilon}_{u_i} \epsilon_{u_{i+1}} \! + \! \bar{\epsilon}_{u_i} \bar{\epsilon}^2_{u_{i+1}} \epsilon_{u_{i+2}} \!\! + \! \ldots + \bar{\epsilon}_{u_i} \bar{\epsilon}^2_{u_{i+1}} \cdots \bar{\epsilon}_{u_{N \!-\! 1}}^2 \epsilon_{u_N}  \\
&= \bar{\epsilon}_{u_i} \sum_{j=i+1}^{N} \{ \epsilon_{u_j} \prod_{k=i+1}^{j - 1} \bar{\epsilon}_{u_k}^2 \}.
\end{align}
Having $P_{c}^{i,i-1}$, the equivalent uplink erasure rate of $x_{i}$ in $(i - 1)$th equation is
\begin{equation}
\epsilon_{u}^{i,i-1} = \epsilon_{u_{i}} ( 1 - P_{c}^{i,i-1}).
\end{equation}

Now, assume that $x_i$ is erased in $i$th equation. It can be found if: i) $x_i$ appears in $(i-1)$th equation while $x_{i-1}$ is erased, ii) Both $x_i$ and $x_{i-1}$ appear in $(i - 1)$th equation and only $x_{i - 1}$ is received by relay in $(i - 2)$th equation, and so on. The last possible situation is when $x_1$ is erased in the first equation while $x_2$ exists and none of $x_j$'s in the second to $(i - 1)$th equations is erased. Thus, the probability of erasure correction for $x_i$ at equation $i$ is
\begin{align}\nonumber
P_{c}^{i,i}  & \! = \! \bar{\epsilon}_{u_i} \epsilon_{u_{i-1}} \! + \! \bar{\epsilon}_{u_i} \bar{\epsilon}^2_{u_{i-1}} \epsilon_{u_{i-2}} \! + \! \ldots + \bar{\epsilon}_{u_i} \bar{\epsilon}^2_{u_{i-1}} \cdots \bar{\epsilon}_{u_{2}}^2 \epsilon_{u_1}  \\
& = \bar{\epsilon}_{u_i} \sum_{j=1}^{i-1} \{ \epsilon_{u_j} \prod_{k=j+1}^{i - 1} \bar{\epsilon}_{u_k}^2 \}.
\end{align}
Similarly, the equivalent uplink erasure rate of $u_i$ when it experiences erasure in $i$th uplink transmission is
\begin{equation}
\epsilon_{u}^{i,i} = \epsilon_{u_{i}} ( 1 - P_{c}^{i,i}).
\end{equation}
Notice that $P_c^{1,1} = P_c^{N,N - 1} = 0$. To apply the effect of reconstruction on EEER calculation, we should properly replace $\epsilon_{u_i}$ with either $\epsilon_{u}^{i,i-1}$ or $\epsilon_{u}^{i,i}$. In other words, $x_i$ experiences erasure in the $i$th row of $\textbf{A}_1$ with $\epsilon_{u}^{i,i-1}$ and with $\epsilon_{u}^{i,i}$ in the $(i + 1)$th row.

\rem For a symmetric EMWRC with MPWR, it can be shown that in the limit of $N \rightarrow \infty$, we have
\begin{align}
E(P_c^{i,i-1}) =  \frac{\bar{\epsilon_u}}{1 + \bar{\epsilon_u}},\\
E(P_c^{i,i}) =  \frac{\bar{\epsilon_u}}{1 + \bar{\epsilon_u}},
\end{align}
where $\bar{\epsilon_u} = 1 - \epsilon_u$ and $E(\cdot)$ is the expected value. As a consequence, both $\epsilon_{u}^{i,i-1}$ and $\epsilon_{u}^{i,i}$ approach $\frac{\epsilon_u}{2 - \epsilon_u}$.

\subsection{EEER Calculation for OPPWR}
\textit{OPPWR without Reconstruction:} Consider one round of communication for OPPWR which consists of $N$ pairwise user transmissions. Since for OPPWR, $\textbf{A}$ is a circulant matrix, without loss of generality, we find $\epsilon_{i,1}$ for $i = 2,3,\ldots,N$. Other pairwise EEERs are similarly found by proper circulation of $\epsilon_{i,1}$.

Having $x_1$ (the first row of $\textbf{A}_1$ in (\ref{Eq A_1})), $u_1$ can find $x_i$ either in row $i$  or $i + 1$ of (\ref{Eq linear system}) for $i = 2,3,\ldots,N$. Let us denote the probability of finding $x_i$ in row $i$ and $i + 1$ by $P_1(i)$ and $P_2(i)$ respectively. Thus, the probability of retrieving $x_i$ in $u_1$, $P(i)$, is
\begin{equation} \label{Pn OOPW}
P(i) = P_1(i) + P_2(i) - P_c(i)
\end{equation}
where $P_c(i)$ is the probability of being able to retrieve $x_i$ in both $i$th and $(i+1)$th rows of $\textbf{A}_1$.

$P_1(i)$ is found similar to (\ref{Eq P_1}). Further, due to the cyclic structure of $\textbf{A}$, it can be shown that $P_2(i) = P_1(N - i + 2)$ for $i=2,3,\ldots,N$. Derivation of $P_c(i)$ is also similar to (\ref{Eq P_c}). To calculate $P_c(i)$ in (\ref{Eq P_c}), we should substitute $P_2(i + 1)$ by $P_2(1) = 1$ when $i = N$. This is because $x_1$ appears with $x_N$ for the second time and is always known at $u_1$. Having all terms in (\ref{Pn OOPW}), $\epsilon_{i,1} = 1 - P(i)$. Further, using the circulant structure of $\textbf{A}$, it can be shown that $\epsilon_{i,j} = \epsilon_{i - j + 1,1}$. Having the pairwise EEERs, $\epsilon_f^{\mathrm{OPPWR}} = \displaystyle{\max_{i,j} \epsilon_{i,j}}$ and users' average erasure rate, $\epsilon_{\mathrm{ave}}^{\mathrm{OPPWR}}$, is simply calculated similar to (\ref{Eq Average erasure MPWR}).

\rem \label{Remark OPPWR} For a symmetric EMWRC, pairwise EEERs are not equal when OPPWR is used. In this case, it can be shown that $\epsilon_f^{\mathrm{OPPWR}} = \epsilon_{\lfloor N/2 \rfloor + 1,1}$.

\textit{OPPWR with Reconstruction:}
Similar to MPWR, we calculate $\epsilon_{u}^{i,i-1}$ and $\epsilon_{u}^{i,i}$ to derive the uplink equivalent erasure rate. With a similar logic, it can be shown that for OPW
\begin{align}\nonumber
P_c^{i,i-1} \! & = \! \bar{\epsilon}_{u_i} \epsilon_{u_{i+1}} \! + \! \bar{\epsilon}_{u_i} \bar{\epsilon}^2_{u_{i+1}} \epsilon_{u_{i+2}} \! + \! \ldots \! \\ \nonumber
& \,\,\,\, + \! \bar{\epsilon}_{u_i} \bar{\epsilon}^2_{u_{i+1}} \cdots \bar{\epsilon}_{u_{N}}^2 \bar{\epsilon}_{u_{1}}^2 \cdots \bar{\epsilon}_{u_{i-2}}^2 \epsilon_{u_{i-1}} \\
& = \bar{\epsilon}_{u_i} \sum_{j=i}^{N + i - 2} \{ \epsilon_{u_{m(j) + 1}} \prod_{k=i}^{j - 1} \bar{\epsilon}_{u_{m(k) + 1}}^2 \}.
\end{align}
and
\begin{align}\nonumber
P_c^{i,i} \! &= \! \bar{\epsilon}_{u_i} \epsilon_{u_{i-1}} \! + \! \bar{\epsilon}_{u_i} \bar{\epsilon}^2_{u_{i-1}} \epsilon_{u_{i-2}} \! + \! \ldots \\ \nonumber
& \,\,\,\, + \bar{\epsilon}_{u_i} \bar{\epsilon}^2_{u_{i-1}} \cdots \bar{\epsilon}_{u_{1}}^2 \bar{\epsilon}_{u_{N}}^2 \cdots \bar{\epsilon}_{u_{i+2}}^2 \epsilon_{u_{i+1}} \\
&= \bar{\epsilon}_{u_i} \sum_{j=1}^{N - 1} \{ \epsilon_{u_{m(i - j)}} \prod_{k=1}^{j - 1} \bar{\epsilon}_{u_{m(i - k)}}^2 \}
\end{align}
where $m(\cdot)$ represents modulo-$N$ operation. Other stages of EEER calculation are similar to what described for MPWR.

\rem For a symmetric EMWRC with OPPWR, it can be shown that for all $i$, $P_c^{i,i-1} = P_c^{i,i} = P_c$. Further, in the limit of  $N \rightarrow \infty$,
\begin{equation}
P_c = \frac{\bar{\epsilon_u}}{1 + \bar{\epsilon_u}}.
\end{equation}
As a consequence, similar to MPWR, $\epsilon_u^{i-1,i} = \epsilon_u^{i,i} = \frac{\epsilon_u}{2 - \epsilon_u}.$

\subsection{Numerical Examples} \label{Subsection Numerical Examples}
Here, we present some numerical examples for EEER of proposed schemes. Further, we discuss how EEER can be decreased by modifying the users' transmission scheduling and employing a shuffled transmission schedule for users. The following cases are for a symmetric EMWRC with uplink and downlink erasure probabilities $\epsilon_u$ and $\epsilon_d$ respectively.

\begin{figure}[h!]
\centering
\includegraphics[width = \columnwidth]{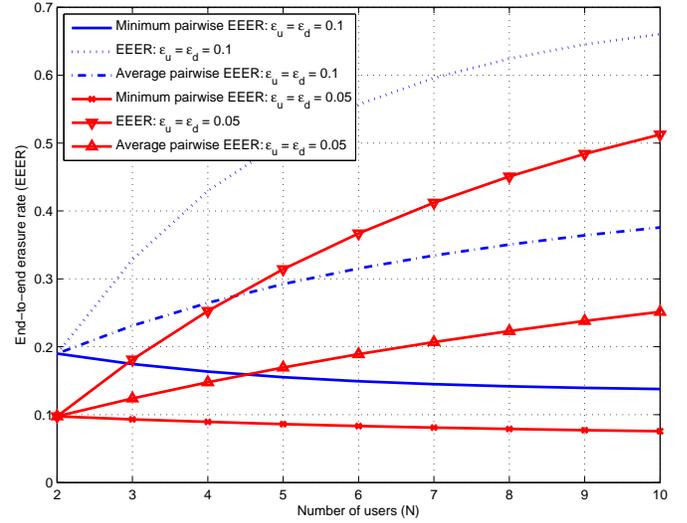}
\caption{EEER, average pairwise EEER and minimum pairwise EEER for MPWR.} \label{Fig EEER for MPWR}
\end{figure}

\begin{figure}[h!]
\centering
\includegraphics[width = \columnwidth]{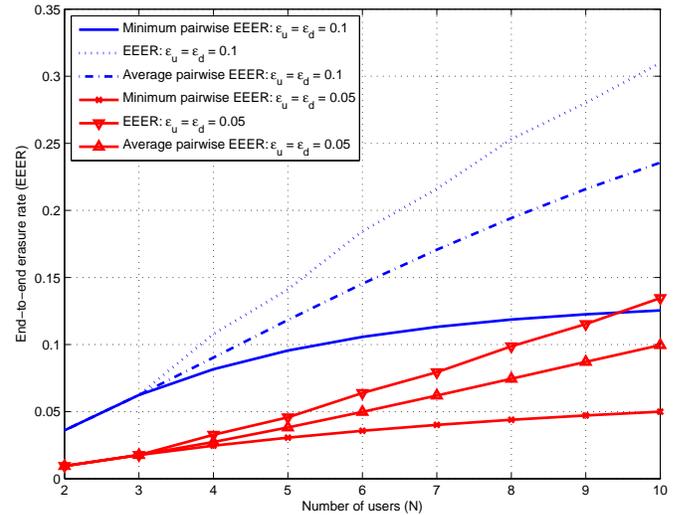}
\caption{EEER, average pairwise EEER and minimum pairwise EEER for OPPWR.} \label{Fig EEER for OPPWR}
\end{figure}

Figure \ref{Fig EEER for MPWR} depicts EEER (maximum pairwise EEER), average pairwise EEER and the minimum pairwise EEER among the users when MPWR is used for two cases when $\epsilon_u = \epsilon_d = 0.1$ and \mbox{$\epsilon_u = \epsilon_d = 0.05$}. As seen, there is a significantly large gap between EEER and average pairwise EEER. Similar results are presented in Figure \ref{Fig EEER for OPPWR} when OPPWR is used. Having such a large variance between pairwise EEERs noticeably limits the achievable rate of the system. These gaps show an unequal pairwise EEERs seen at an arbitrary user (See Remark \ref{Remark MPWR} and \ref{Remark OPPWR}).  Unequal pairwise EEERs appear due to the dependency between the data separation of different users which in turn may result in the erasure propagation in the system. The erasure propagation refers to the situations where one erasure in the uplink or downlink may result in more than one packet erasure after data separation in the users. For example, consider a four-user scenario with MPWR where the uplink transmissions are $x_1 \oplus x_2$, $x_2 \oplus x_3$, and $x_3 \oplus x_4$. If the second equation, i.e. $x_2 \oplus x_3$, is erased in the downlink, both $x_3$ and $x_4$ are lost at $u_1$. It means that one downlink erasure causes two packet erasures. Please note that for OWR, all pairwise EEERs are equal, thus, numerical results are omitted here. 

To improve the system's achievable rate, it is desired to decrease EEER by evening out the pairwise EEERs at the users. For this purpose, we suggest using a shuffled (random) transmission scheduling which evenly spreads the erasure propagation over all users and narrows the gap between the EEER and the average pairwise EEER. In this approach, all users have psuedorandom number generators with the same initial seeds. Thus, the output of number generators are equal at all users. For each round of communication, psuedorandom number generators give a random permutation of numbers from 1 to $N$. We denote this psuedorandom sequence by $\{S_1,S_2,\ldots,S_N \}$. This random sequence specifies the order of transmission by users. For our proposed pairwise schemes, in the first uplink transmission, user $S_1$ and user $S_2$ transmit, in the second uplink transmission, user $S_2$ and user $S_3$ transmit and so on. For OPPWR, user $S_N$ and user $S_1$ also transmit together in the last uplink slot.

In the abovementioned shuffled scheduling, $i$th row of $\textbf{A}$ is assigned to the pairwise transmission of $u_{S_i}$ and $u_{S_{i + 1}}$ for each round of communication. Note that $u_{S_i}$ and $u_{S_{i + 1}}$ can be any arbitrary two users from $u_1$ to $u_N$ in each round. Thus, by doing shuffled scheduling over large number of communication rounds, we expect EEER and minimum pairwise EEER to converge to the average pairwise EEER. As a consequence, shuffled transmission scheduling significantly evens out the pairwise erasure rates resulting in a lower overall EEER.

\begin{figure}[h!]
\centering
\includegraphics[width = \columnwidth]{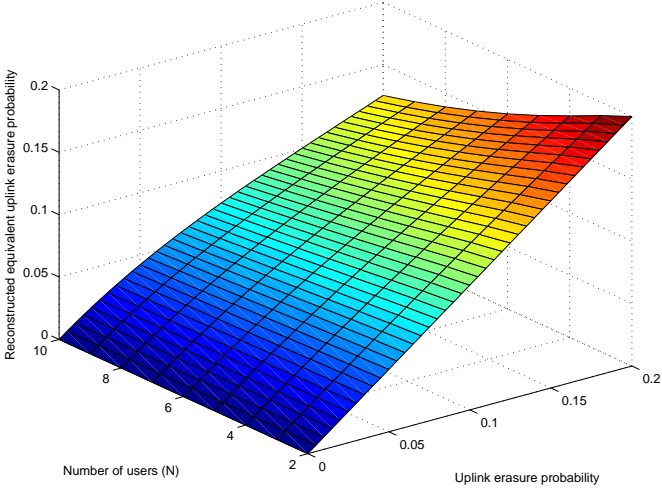}
\caption{Equivalent Uplink erasure probability for MPWR.} \label{Fig Equivalent Uplink erasure for MPWR}
\end{figure}

\begin{figure}[h!]
\centering
\includegraphics[width = \columnwidth]{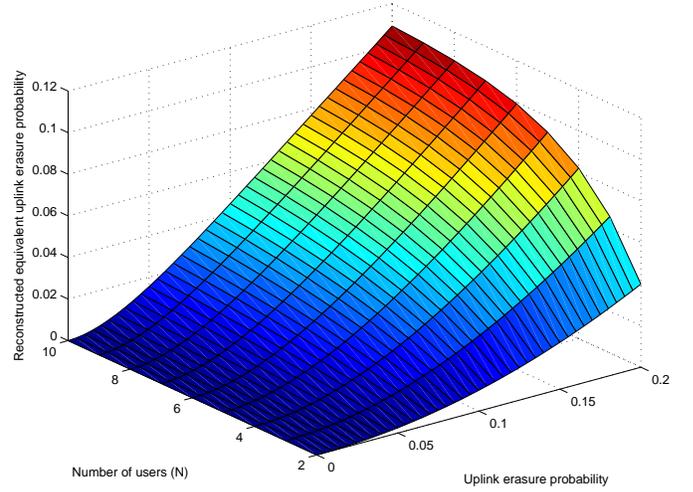}
\caption{Equivalent Uplink erasure for probability OPPWR.} \label{Fig Equivalent Uplink erasure for OPPWR}
\end{figure}

Effect of the reconstruction on the equivalent uplink erasure probability is presented in Figure~\ref{Fig Equivalent Uplink erasure for MPWR} and Figure~\ref{Fig Equivalent Uplink erasure for OPPWR} for MPWR and OPPWR, respectively. In these figures, the average equivalent uplink erasure probability over all users is depicted versus the uplink erasure probability and the number of users. As seen, for small $N$, reconstruction is not much helpful when MPWR is used. For instance, if $N = 2$, reconstruction does not improve the performance at all since the data of each user (here, two users) exist in only one uplink transmission. Hence, there is no redundancy for retrieving the users' data from other uplink transmissions if it is erased. On the other hand, reconstruction causes the best improvement in terms of erasure rate for OPPWR when $N = 2$. This is due to the repetitive transmission of users' data (each user's data packet is sent twice). As number of users increases, performance improvement by reconstruction increases for MPWR while it decreases for OPPWR. However, generally speaking, reconstruction at the relay has a more significant improvement for OPPWR.

\section{Rate Upper bound} \label{Section Upper bound}
In this section, we derive an upper bound on the achievable common data rate, $R$, for the described EMWRC. This bound is later used to evaluate the performance of the proposed data-sharing schemes. To find the rate bound, we apply cut-set theorem \cite{Cover_Relay_TIT}.

\begin{figure}[h!]
\centering
\psfrag{u1}{$u_1$}
\psfrag{u2}{$u_2$}
\psfrag{u3}{$u_{i - 1}$}
\psfrag{u4}{$u_{i + 1}$}
\psfrag{u5}{$u_{N - 1}$}
\psfrag{u6}{$u_{N}$}
\psfrag{u7}{$u_{i}$}
\includegraphics[scale =.19] {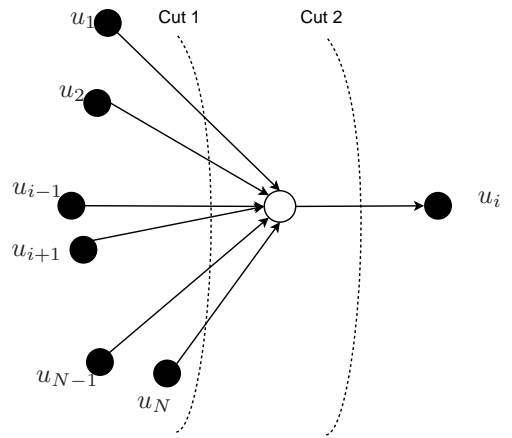}
\caption{Cut-sets used to find the rate upper bound} \label{Fig Cut_set}
\end{figure}

To start, we first consider data transmission from other users to $u_i$ and derive the rate upper bound in this case. For this user, two cuts are considered (Figure \ref{Fig Cut_set}): the cut considering the relay and $u_i$ as receivers of a multiple-access channel interested in decoding the data of other $N\!-\!1$ users, and the cut considering the relay as the transmitter to $u_i$. For the first cut, the data rate is limited by the user with the worst uplink erasure rate as well as the sum-rate condition. Using similar arguments as \cite{Vishwanath_ITW_07}, it is easy to show that the sum-rate for the first cut is bounded by $1\!-\!\prod_{j =1,j \neq i}^{N} \epsilon_{u_j}$. Thus, by denoting the transmitted common data rate from other users to $u_i$ by $R_i$, we have
\begin{align}
R_i &\leq \min \{ \min_{j =1,j \neq i} \{1 - \epsilon_{u_i}\}, \frac{1}{(N - 1)} (1\!-\!\prod_{j =1,j \neq i}^{N} \epsilon_{u_j} ) \}.
\end{align}
The second cut is a simple single user erasure channel. Thus,
\begin{equation}
R_i \le  \frac{1}{N - 1} (1-\epsilon_{d_i}).
\end{equation}
Now, if we repeat the cut-set discussion for all $u_i$'s, the achievable common rate is $R = \displaystyle{\min_i R_i}$.

\section{Performance Analysis} \label{Section Performance}
In this section, we study the performance of the three aformentioned schemes (i.e. OWR, MPWR and OPPWR) in terms of their achievable rate and the transmission overhead for the data exchange between the users. Here, we assume a symmetric EMWRC with uplink and downlink erasure probabilities $\epsilon_u$ and $\epsilon_d$.

The achievable rate of the schemes is determined by the worst erasure rate between a pair of users which is reflected in EEER. In addition to EEER, the number of consumed uplink and downlink slots (number of channel uses) for data exchange between users is also important for to make a fair comparison between the schemes. To this end, we consider the normalized achievable rate which is the carried data over one uplink and downlink time slots. According to this definition, the normalized achievable rate for OWR, MPWR and OPPWR are $R_{\mathrm{OWR}} = (1 - \epsilon_f^{\mathrm{OWR}})/ N$, $R_{\mathrm{MPWR}} = (1 - \epsilon_{f}^{\mathrm{MPWR}})/ (N - 1)$ and $R_{\mathrm{OPPWR}} = (1 - \epsilon_{f}^{\mathrm{OPPWR}}) /N$ respectively.

\begin{figure}[h!]
\centering
\includegraphics[width = \columnwidth]{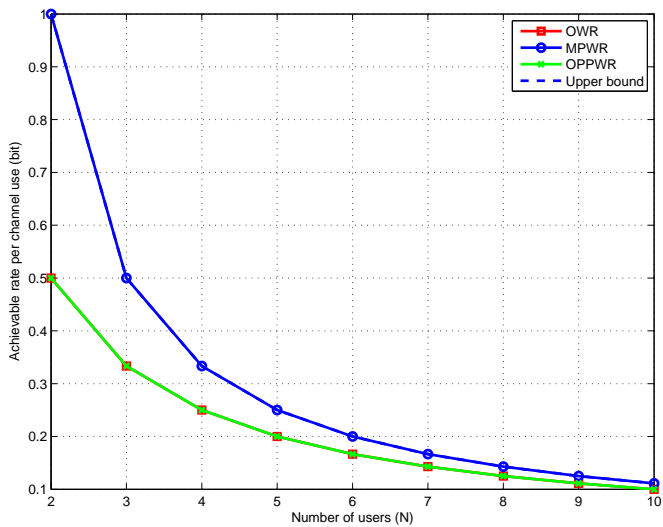}
\caption{Achievable data rates and the rate upper bound when  $\epsilon_u = \epsilon_d = 0$.} \label{Fig rate cut-set 1}
\end{figure}

Figure~\ref{Fig rate cut-set 1} depicts the comparison between the normalized achievable rates of OWR, MPWR, OPPWR, and the rate upper bound (derived in Section \ref{Section System model}) for an ideal channel with no erasure, i.e $\epsilon_d = \epsilon_u = 0$. As seen, MPWR can actually achieve the upper bound for such an ideal channel since its division factor, $N - 1$, is equal to the division factor of the upper bound. Also, OPPWR and OWR provide equal rates which always fall under the upper bound and the achievable rates of MPWR. By increasing the number of users, the superiority of MPWR over OPPWR and OWR decreases since the ratio of their division factors, i.e. $\frac{N - 1}{N}$ decreases.

\begin{figure}[h!]
\centering
\includegraphics[width = \columnwidth]{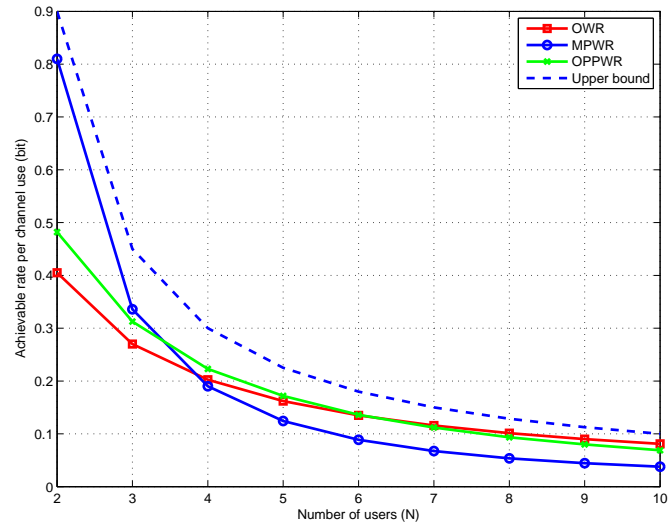}
\caption{Achievable data rates and the rate upper bound when $\epsilon_u = \epsilon_d = 0.1$.} \label{Fig rate cut-set 2 without}
\end{figure}

\begin{figure}[h!]
\centering
\includegraphics[width = \columnwidth]{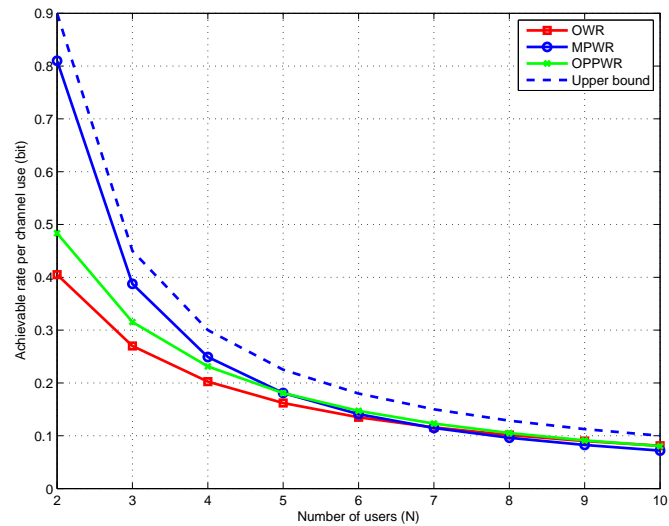}
\caption{Achievable data rates and the rate upper bound when $\epsilon_u = \epsilon_d = 0.1$ and reconstruction and shuffled scheduling are applied.} \label{Fig rate cut-set 2 with}
\end{figure}

By increasing the erasure rate of channels, MPWR is no longer the best approach. The results are shown for a more realistic channel with $\epsilon_u = 0.1$ and $\epsilon_d = 0.1$ in Figure~\ref{Fig rate cut-set 2 without}. As seen, for $N \leq 4$, $5 \leq N \leq 8$, and $9 \leq N$, MPWR, OPPWR, and OWR achieve the highest normalized rate. To investigate the effect of reconstruction at the relay as well as the shuffled transmission scheduling, numerical results for symmetric channels with $\epsilon_u = \epsilon_d = 0.1$ are presented in Figure~\ref{Fig rate cut-set 2 with}. Using reconstruction and shuffled scheduling improves the achievable rates of proposed pairwise schemes, specially MPWR. Note that MPWR and OPPWR may suffer from erasure propagation events. This can be the reason for smaller gains from MPWR and OPPWR when $N$ increases.

\begin{figure}[h!]
\centering
\includegraphics[width = \columnwidth]{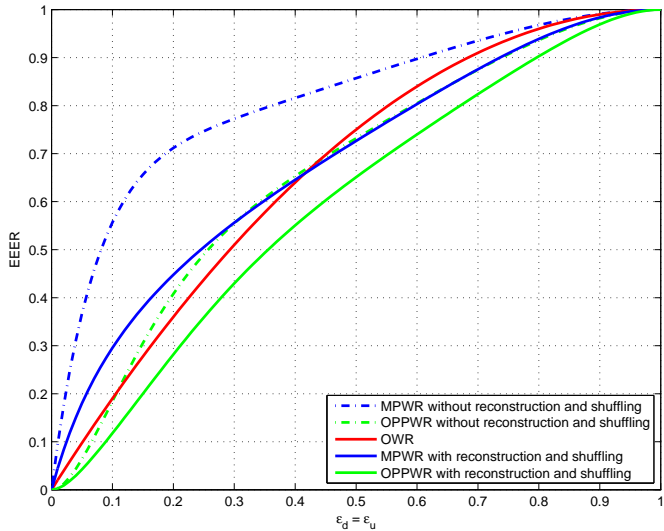}
\caption{EEER comparison under different scenarios when $N = 6$.} \label{Fig EEER comparison}
\end{figure}

To better illustrate the performance improvement of random shuffling and relay reconstruction, a comparison between EEER for MPWR, OPPWR and OWR is presented in Figure \ref{Fig EEER comparison} when \mbox{$N = 6$}. Without reconstruction or shuffled transmission, EEER of OWR resides under the EEER of MPWR. However, using these two techniques significantly reduces MPWR's EEER and for some erasure probabilities, MPWR's EEER is less than OWR's EEER. Similar behavior is observed for OPPWR where using reconstruction and shuffled scheduling results in outperforming OWR by OPPWR over all erasure probabilities.

\begin{figure}[h!]
\centering
\includegraphics[width = \columnwidth]{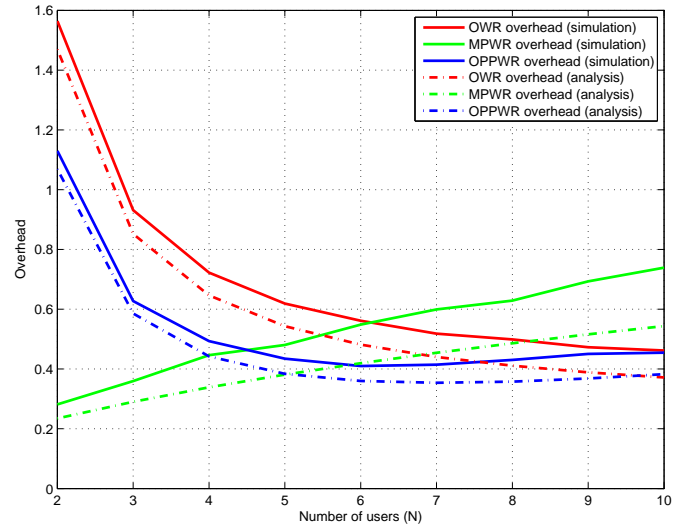}
\caption{Overhead comparison for $\epsilon_u=0.1, \epsilon_d=0.1 $.} \label{Fig Overhead}
\end{figure}

Figure~\ref{Fig Overhead} depicts the simulation and analytical results for the transmission overhead of different schemes when $\epsilon_u= \epsilon_d=0.1$. Similar to the achievable rates, here, the transmission overhead for different schemes are normalized. Transmission overhead can be considered as a notion of delay in EMWRC. More specifically, if the normalized overhead is $O_N$ and the transmission time for each packet is $T_p$ seconds, then the whole communication process takes $(1 + O_N) K T_p$ seconds to accomplish. This means that smaller overhead gives shorter transmission time and consequently smaller communication delay.

For simulation, a Raptor code with information length 14000 and an outer code (LDPC) of rate 0.9872 has been used for fountain coding. The degree distribution for the considered fountain code is derived based on \cite{Sanghavi_LT_ITW2007}. To this end, we define the following variables
\begin{align} 
&Q(1) = 0.01, \\ \nonumber
&Q(i) = \frac{1}{ai(i - 1)} \quad 2 \leq i \leq m - 1, \\ \nonumber
&Q(m) = 1 - \frac{m - 2}{a(m - 1)},\\ \nonumber
&Q(i) = 0 \quad \text{Otherwise},
\end{align}
where $m = 78$ based on our code rate, $a$ is 
\begin{equation}
a = \frac{m - 1}{m} + \frac{1}{m z^{m - 1}} \sum_{i \geq m} \frac{z^i}{i}
\end{equation}
and $z = 0.9872$ is the code rate. Now, the probability of degree $i$ in the code is defined as 
\begin{equation}
P(i) = \frac{Q(i)}{\displaystyle{\sum_{i = 1}^m Q(i)}}.
\end{equation}

Also, in the simulation setup, a shuffled transmission schedule is used and relay performs reconstruction to reduce the effective uplink erasure rate. The analytical results are calculated  using EEER as explained in Section~\ref{Section EEER}. Note that there is a gap between the analytical and simulation results due to assuming ideal fountain code in the analytical overhead calculation. However, using EEER, the overhead of the schemes can be evaluated well without the need for tedious computer simulations.

\section{Conclusion} \label{Section Conclusion}
In this paper, we studied low-latency data sharing schemes for EMWRCs. To this end, we first mentioned the challenges confronting the use of fountain coding for EMWRCs. Then, we proposed two simple low-latency data sharing schemes, namely MPWR and OPPWR, based on fountain coding. We also showed that by performing simple matrix operations at the relay and shuffling the order of users' transmissions, the performance of MPWR and OPPWR can be further enhanced. To find the achievable data rate and transmission overhead of our solutions, we introduced EEER and calculated it analytically for our strategies. In addition, an upper bound on the achievable rate of EMWRCs was derived. The achievable rates of MPWR and OPPWR were then compared with this bound as well as the achievable rates of OWR. This comparison along with comparing the transmission overhead of MPWR, OPPWR and OWR revealed that for small $N$, MPWR has the best performance. By increasing $N$, first OPPWR and then OWR outperform the other two schemes. Seeking methods to improve the performance of data sharing schemes over EMWRCs, for instance through smarter users' and relay transmission strategies, is considered to future research directions.  Furthermore, application of rateless coding for multi-way relaying over wireless fading environment can be explored  as an extension of this work. This may demand the modification of the users' and relay's transmission strategies according to characteristics of the wireless links.

\section{Acknowledgments}
The authors would like to appreciate Kaveh Mahdaviani for providing us some parts of simulation codes. Further, we thank Natural Sciences and Engineering Research Council of Canada (NSERC) and Alberta Innovates Technology Futures (AITF) for supporting our research.

\bibliographystyle{IEEEtran}
\bibliography{IEEEabrv,moslembib}

\end{document}